\documentclass[aps,prb,twocolumn,showpacs,floatfix,a4paper]{revtex4}
\usepackage{graphicx}
\usepackage[intlimits]{amsmath}
\usepackage{color}

\begin{document}

\title{On low eigevalues of the entanglement Hamiltonian, localization length, and rare regions
in disordered interacting one-dimensional systems}
\author{Richard Berkovits}
\affiliation{Department of Physics, Jack and Pearl Resnick Institute, Bar-Ilan University, Ramat-Gan 52900, Israel}

\begin{abstract}
  The properties of the low-lying eigenvalues of the entanglement Hamiltonian and
  their relation to the localization length of a disordered interacting one-dimensional
  many-particle system is studied.
  The average of the first entanglement Hamiltonian level spacing is proportional to
the ground state
localization length and shows the same dependence on the disorder
and interaction strength as the localization length. This is the result of
the fact that entanglement is limited to distances of order of 
the localization length. The distribution
of the first entanglement level spacing shows 
a Gaussian-like behavior as expected
for level spacings much larger than the disorder broadening.  
For weakly disordered systems
(localization length larger than sample length), the distribution shows
an additional peak at low level spacings. This stems from
rare regions in some samples which exhibit metallic-like behavior
of large entanglement and large particle number fluctuations.
These intermediate “microemulsion” metallic regions embedded
in the insulating phase are discussed. 

\end{abstract}


\maketitle

\section{Introduction}
Close to 60 years after the concept of localization has been introduced 
by Anderson\cite{anderson58},
the localization transition 
remains at the center of many current topics, from applications of
many body in quantum
information\cite{choi16} to random lasing\cite{stano13}.
While for most quantum phases
the gap between the ground state energy and the first excited state
defines the correlation length,
in the localized phase the correlation length corresponds to
the localization length, determined by the exponential
dependence of the conductance, $G$, on the linear
dimension of the system, $L$. The localization length, $\xi$, 
is defined
through the exponential decrease in the conductance 
in the localized phase
$G(L) \sim \exp(-L/\xi)$ \cite{lee85}.

\begin{figure}
\includegraphics[width=8cm,height=!]{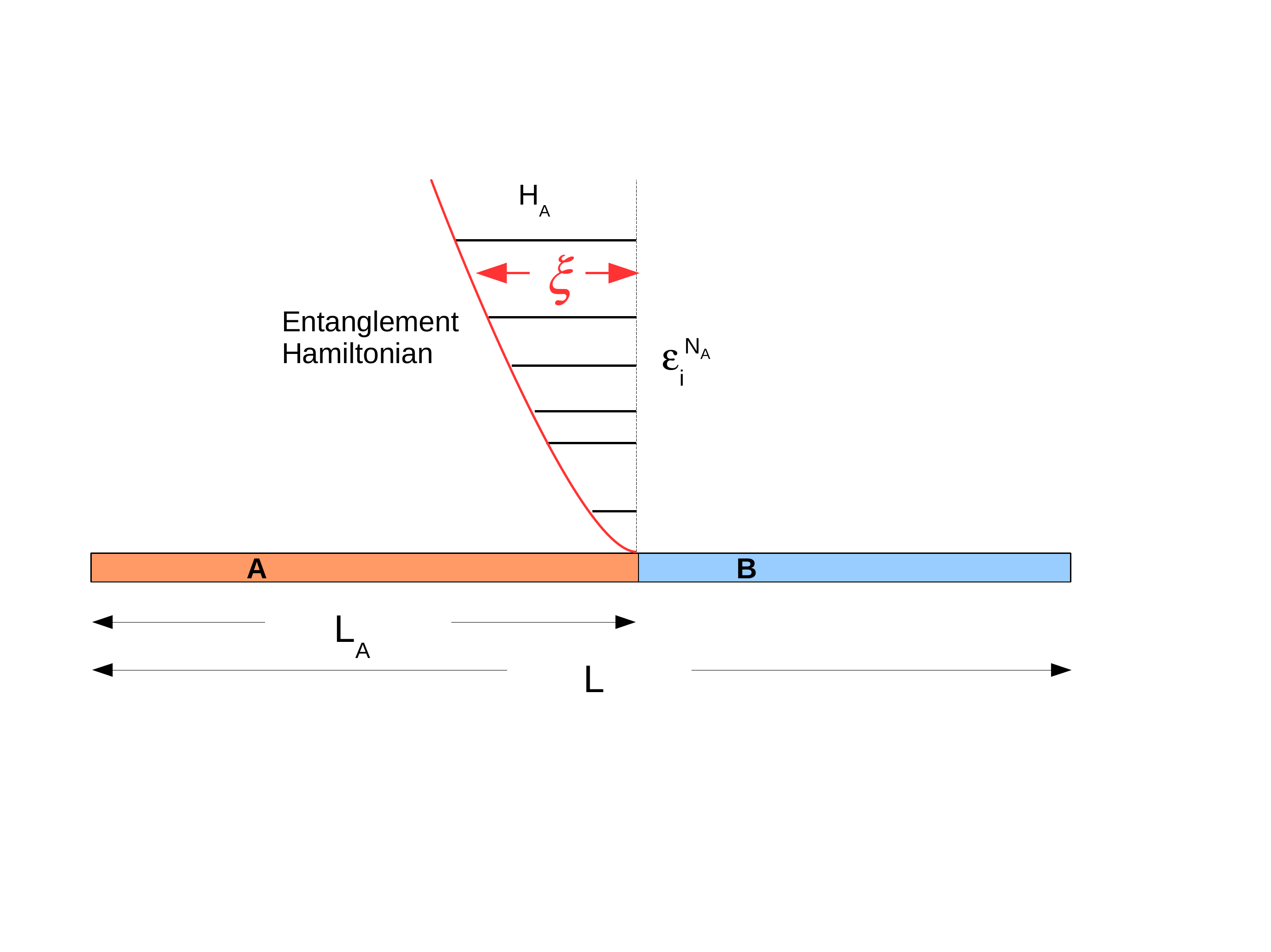}
\caption{\label{fig0}
A schematic representation of the
system and the entanglement Hamiltonian $H_A$. 
A one-dimensional system of length $L$ is bisected into two regions
A and B, with length $L_A$ and $L-L_A$. The reduced density matrix of region
A, $\rho_A$ is calculated and its eigenvalues $\lambda_i^{N_A}$ are used
to construct the entanglement Hamiltonian $H_A$ and its eigenvalues
$\varepsilon_i^{N_A}$. $H_A$ samples the behavior of the system
on a scale $\xi$ from the A-B boundary}
\end{figure}

Recently, it was realized that $\xi$ should also
determine the entanglement properties of a system in the
strongly localized regime \cite{berkovits12}. One does not expect regions
beyond the distance $\xi$ to be entangled.
Thus, by
dividing a one dimensional system into two regions (see Fig. \ref{fig0}),
and studying the entanglement between them, one may hope to gain a measure
of $\xi$ through the entanglement properties.
Indeed, the averaged entanglement entropy increases logarithmically as
long as the length of region A, $L_A$ is smaller than 
than $\xi$ and saturates for $L_A>\xi$ \cite{berkovits12,chu13}. 
Here we would like to use entanglement as a window
into the physics of the region within length $\xi$ 
from the boundary. 

The information regarding the entanglement between the regions A and B
is encoded in
the reduced density matrix (RDM), $\rho_{A(B)}$, 
of regions A (or B).
For a system in a pure state $|\Psi\rangle$, 
$\rho_A$ is defined as:
$\rho_{A}={\rm Tr}_{B}|\Psi\rangle\langle \Psi |$, where the
degrees of freedom of region B are traced out.
It is important to note that diagonalizing $\rho_A$
defines a basis that completely spans the Hilbert space of region A and
if there exists any conserved quantum number (for the
case discussed in this paper the conserved quantum number is $N_A$,
the number of particles
in region A) that basis is also the eigenvectors of $N_A$.
The eigenvalues of the RDM, $\lambda_i^{N_A}$, are used to
extract measures for the entanglement between the regions, such as the
entanglement entropy, defined as:
$S_{A}=- \sum_i \lambda_i^{N_A} \ln \lambda_i^{N_AA}$,
and the R\'enyi entropy:
$S_{n A}=- \frac{1}{1-n} \ln \sum_i (\lambda_i^{N_A})^n$.
where the first R\'enyi entropy ($n \rightarrow 1$) is equal to the entanglement entropy.

Recently Li and Haldane \cite{haldane08} have suggested a different
way to interpret the eigenvalues of the RDM.
They noted that the RDM of region A may be seen as
a density matrix of a mixed thermal state of an ersatz system
described by a Hamiltonian $H_A$ such that
$\rho_A=\exp(-\beta H_A)$, where $H_A$ is known as the entanglement Hamiltonian,
and $\beta=1$. Under these conditions the 
eigenvalues  of $H_A$ are given by
$\varepsilon^{N_A}_i=-\ln(\lambda^{N_A}_i)$.
Up until know, these are just mathematical manipulations, but Li and Haldane
noted that for a fractional quantum Hall 
$\nu=5/2$ state where the edge and the bulk are
chosen as the two regions, the eigenvalues $\varepsilon^{N_A}_i$,
resembled the true edge
excitation spectrum. This is actually quite intuitive since
due to geometry, the boundary between regions A and B and the
volume of region A are the same, and therefore one should expect
$H_A$ to give a reasonable representation of the physics
of region A.
This suggests
that the low-energy entanglement Hamiltonian spectrum
shows some correspondence to the true many-body
excitations of the partitioned segment (the edge region).
This correspondence has been demonstrated for different 
cases of topological insulators \cite{laflorencie16,koch17}.

The opposite situation is the entanglement between regions A and B
for a 1D system depicted in Fig. \ref{fig0} where the contact between
the two regions is a point. As pointed out by Alba et. al. \cite{alba12},
for different 1D gaped systems (i.e., systems with finite correlation length)
one expects that the entanglement spectrum will be mainly influenced
by a region of order of the correlation length from the boundary.
This is similar to the situation in the Anderson localized phase, where
there is no gap but the localization length plays the role of a correlation
length \cite{lee85}. This is highlighted by the saturation of the entanglement
entropy once $L_A>\xi$ discussed earlier, as well as by power-law behavior
which depends on the localization length
of the entanglement spectrum of a highly excited state\cite{serbyn16}
in the many-body localized regime.

In this paper we show that the entanglement spectrum
of the ground and low-lying excitation
states of a localized interacting 1D many-particle system
shows a clear signature of the many-particle localization length.
Specifically, the first level spacing of the entanglement 
energies for a given $N_A$,
$\Delta^{N_A}_1=\varepsilon^{N_A}_2-\varepsilon^{N_A}_1$, is proportional
to $1/\sqrt{\xi}$. As we shall explain this is the behavior expected
from a many particle state trapped in a potential of width
$\xi$ from the boundary. Moreover, the proportionality depends
linearly on the strength of particle-particle interactions, $U$, as expected
from charge particles trapped in a potential. It is also shown that 
deep in the localized regime, where no difference in the localization
length between the many-particle
ground state and the low-lying states is expected, also
$\Delta^{N_A}_1$ is similar. On the other hand, for weaker disorder, where
the localization length for low-lying excitations is significantly
larger than for the ground state \cite{berkovits14}, also 
$\Delta^{N_A}_1$ becomes smaller.
The distribution of $\Delta^{N_A}_1$ becomes Gaussian for $\xi \ll L_A$,
with a width proportional to disorder and does not depend on $U$, as might
be expected for the level spacing of states in a weakly disordered
quantum dot.

This is important not only as a new simple way if determining the localization
length of an interacting many-particle system, but mainly as a way to
access the properties of disordered many-particle systems on short
length scales (of scale $\xi$) and large energy scales $U/\sqrt{\xi}$. This is
illustrated by using the low lying values of $\varepsilon$ to detect and
characterize rare regions in the sample which appear for low disorder strongly
interacting samples. These regions exhibit metallic-like
behavior such as large entanglement and high particle number variance.

The paper is organized as follows: In Sec. \ref{s1} the model for the
interacting fermions in a disordered one dimensional system is defined.
The next section (Sec. \ref{s2}) discusses the average of the
first entanglement level spacing and its relation to the ground state
localization length. The following section (Sec. \ref{s3})
investigates the properties of the distribution
of the first entanglement level spacing. The appearance of some
rare regions in the sample which exhibit metallic-like features is
discussed in Sec. \ref{s4}. The significance of the results and relevance
to further work is discussed if Sec. \ref{s5}.

\section{Model}
\label{s1}

In this paper
we study spinless electrons confined to a 1D wire of
length $L$ 
with repulsive nearest-neighbor particle-particle
interactions and on-site disordered potential, depicted
by the Hamiltonian:
\begin{eqnarray} \label{hamiltonian}
H &=& 
\displaystyle \sum_{j=1}^{L} \epsilon_j {\hat c}^{\dagger}_{j}{\hat c}_{j}
-t \displaystyle \sum_{j=1}^{L-1}({\hat c}^{\dagger}_{j}{\hat c}_{j+1} + h.c.) \\ \nonumber
&+& U \displaystyle \sum_{j=1}^{L-1}({\hat c}^{\dagger}_{j}{\hat c}_{j} - \frac{1}{2})
({\hat c}^{\dagger}_{j+1}{\hat c}_{j+1} - \frac{1}{2}),
\end{eqnarray}
in which $\epsilon_j$ represents the on-site energy
drawn from a uniform 
distribution $[-W/2,W/2]$,
${\hat c}_j^{\dagger}$ is the creation 
operator of a spinless particle at site $j$,
and $t=1$ is the
hopping matrix element.
The interaction strength is $U$, with 
a background positive charge.
For this model the localization
length $\xi_0 \approx 105/W^2$ \cite{romer97} for $U=0$. 
The Luttinger parameter is defined as $K(U)=\pi/[2 \cos ^{-1} (-U/2)]$ 
\cite{g_formula,giamarchi03}, where $K(U=0)=1$. As $U$ increases
$K$ becomes smaller, and at the transition to a charge density
wave ($U=2$) $K=1/2$.
Renormalization group scaling 
of the localization length 
suggests $\xi = (\xi_0)^{1/(3-2K)}$ \cite{apel_82,giamarchi88},
i.e., $\xi$ decreases as the interaction increases and the system becomes
more localized.

\section{Averaged first entanglement level spacing}
\label{s2}

We use the numerical DMRG \cite{white92,dmrg}
method to calculate the RDM for 
the ground state and low-lying excited states of
the Hamiltonian depicted in Eq. (\ref{hamiltonian})
\cite{berkovits12,berkovits14} at half-filling.
We calculate the eigenvalues of the RDM for a system of length $L=700$,
different values of disorder $W=0.3,0.7,1.5,2.5,3.5,4,5$
corresponding to $\xi \sim 1100,200,50,20,9,6.5,4$,
different values of interaction strength $U=0,0.3,0.6,0.9,1.2,1.5,1.8,2.1,2.4$
corresponding to $K=1,0.91,0.84,0.77,0.71,0.65,0.58,0.49,0.48$ (the
two last values are a continuation of the values $K$ for $U=2$)
and different sizes of region A: usually, $L_A=10,20,\ldots,L-10$, for at least $100$
realizations of on-site disorder.

Here we concentrate on the first level spacing of the entanglement levels.
Since the number of electrons in region A remains a good quantum number 
of $H_A$, for each eigenstate of $\rho_A$ one should calculate
both $\lambda_i$ and $N^A_i$ (the number of particles in the region $A$).
Calculating $N^A_i$ does not add to the complexity of 
the DMRG code \cite{song12}.
Thus, the average first level spacing
$\Delta^{N_A}_1(L_A)= \langle
\varepsilon^{N_A}_2(L_A) - \varepsilon^{N_A}_1 (L_A) \rangle$
(where $\langle \ldots \rangle$ depicts an average over different realization of disorder) is calculated.
It is important to note that if one ignores the subscript $N_A$ and calculates
$\varepsilon_{2} - \varepsilon_1$ one gets a very different result. Essentially,
since eigen-states of $\rho_A$ belonging to different sectors of $N_A$ do not couple,
one obtains a spacing of two unrelated energies which does not contain much
physical information. On the other hand
$\Delta^{N_A}_1$ does not depend strongly on $N_A$ and therefore we also average over
values of $N_A \sim L_A/2$ to obtain $\Delta_1$.

\begin{figure}
\includegraphics[width=8cm,height=!]{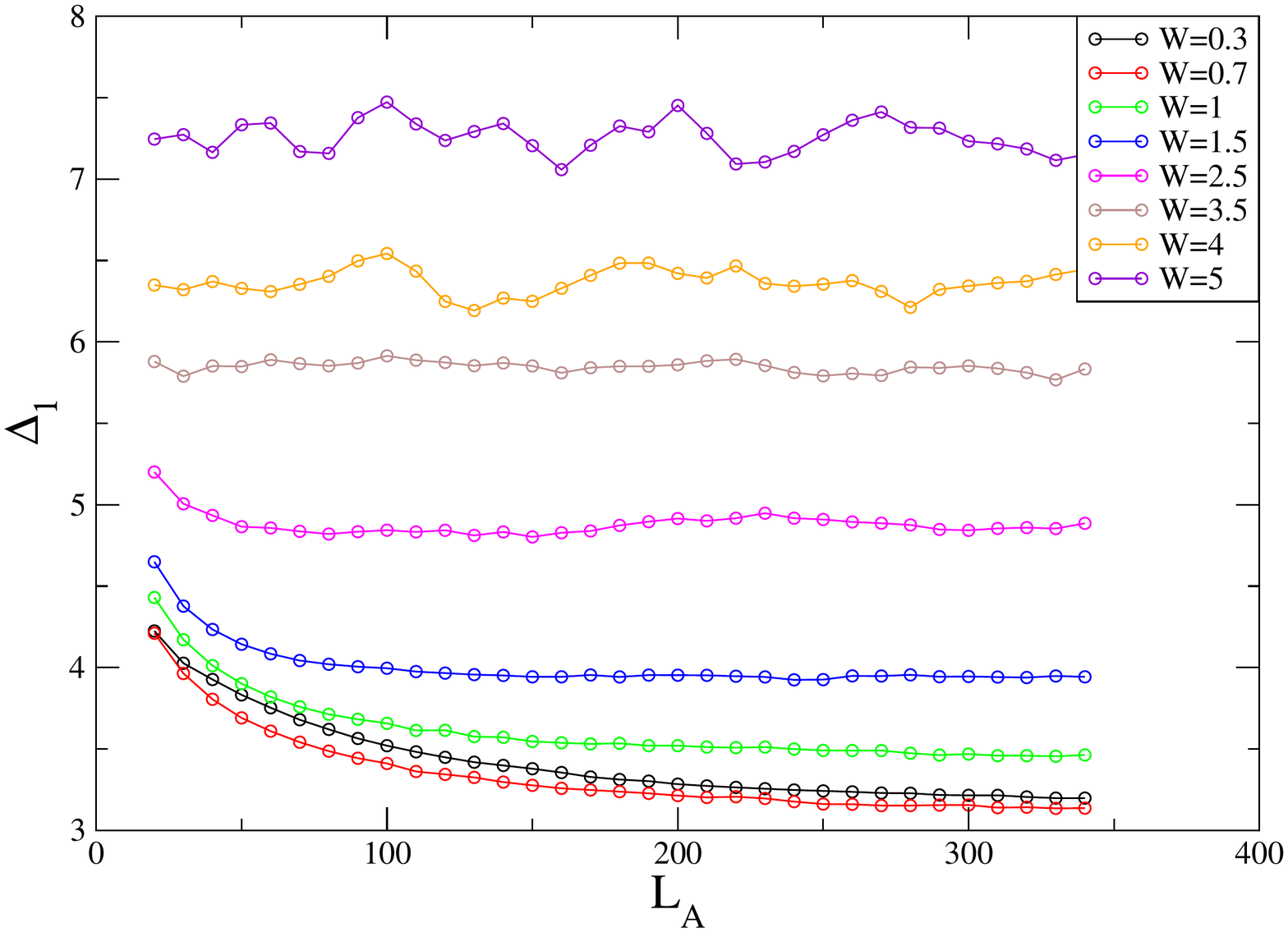}
\includegraphics[width=8cm,height=!]{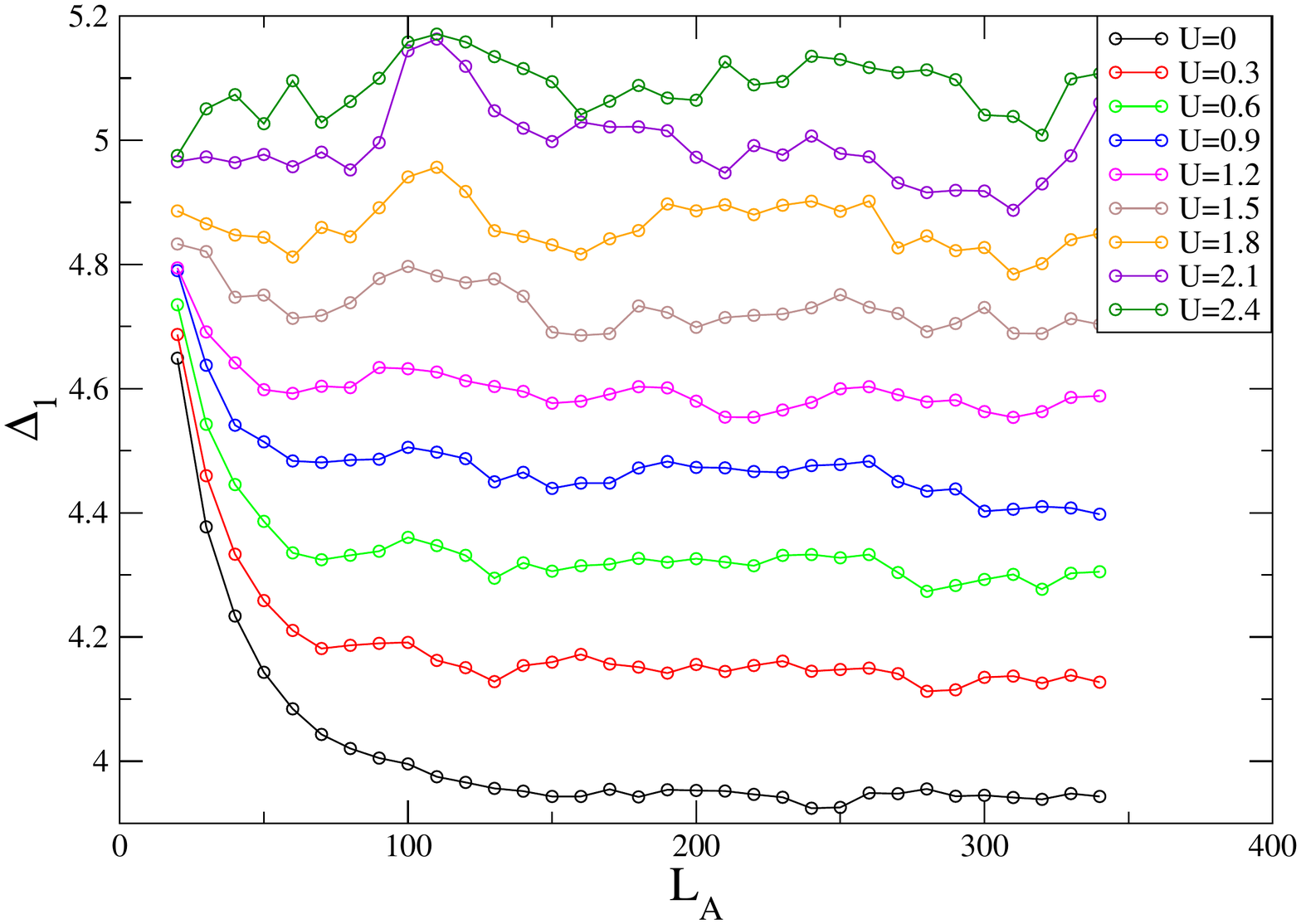}
\caption{\label{fig1}
The averaged first level spacing $\Delta_1$ as function of the size of region $A$,
$L_A$ for a system of size $L=700$ and
different values of disorder $W$ or interaction strength $U$. Top:
The non-interacting case for different values of $W$. Bottom: For $W=1.5$ ($\xi_0=50$)
and different values of $U$. In both cases it is clear that as $\xi$ decreases whether
due to increasing disorder or interaction strength, $\Delta_1$ saturates at smaller
values of $L_A$, indicating a finite region in $A$ influenced by the entanglement.
}
\end{figure}

In Fig. \ref{fig1}, $\Delta_1(L_A)$ for different values of disorder $W$ and interaction strength $U$ is
presented. Since we expect that there will be no entanglement beyond a region proportional
to $\xi$, the entanglement spectrum should be affected by the shortest length scale of $L_A$ or
$\xi$. Thus, $\Delta_1(L_A>\xi) = \Delta_1(\xi)$ should saturate, which is indeed seen
for higher values of $W$ and $U$ for which $\xi$ becomes shorter. Another feature
of $\Delta_1(L_A)$ which should be considered is its magnitude. A simple consideration, treating
the fact that the entanglement is confined to the boundary between the regions as an effective
confining potential of width $\xi$ in $H_A$ (see Fig \ref{fig0}), will result in 
$\Delta_1(\xi) \propto 1/\xi$.
This argument neglects the fact that we are calculating the level spacing within
the same sector $N_A$. Thus one must consider that the next state might not belong to the same
$N_A$ sector. Taking into account that the variance in the number of particles in region A
is proportional to $\sqrt{\xi}$ one concludes that $\Delta_1(\xi) \propto 1/\sqrt{\xi}$.

\begin{figure}
\includegraphics[width=8cm,height=!]{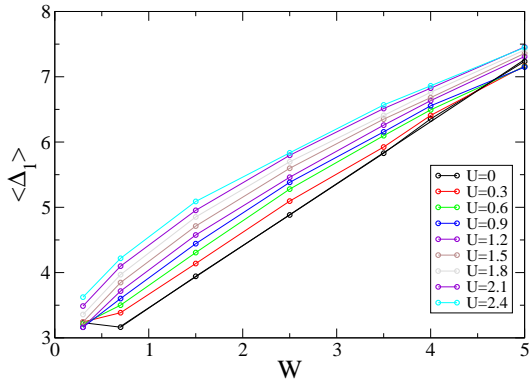}
\includegraphics[width=8cm,height=!]{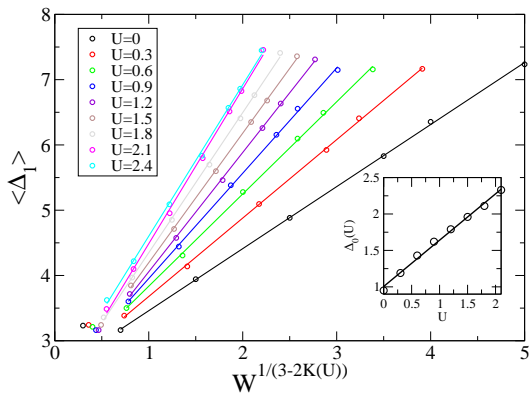}
\caption{\label{fig2}
The averaged first level spacing $\Delta_1$ as function of the disorder $W$ and interaction
strength $U$ averaged also over the central region on the wire $L/4<L_A<3L_A/4$ (symbols). Top:
$\Delta_1$ as function of the disorder strength. For the non-interacting case ($U=0$) as long
as $\xi<L$ (i.e., as long as $W>0.5$)  $\Delta_1$ is linearly dependent on $W$ (black line), in line
with the expectation from  $\Delta_1 \propto 1/\sqrt{\xi_0}$. This relation does not hold for the
interacting case $U>0$. Bottom: Taking into account the influence of $U$ on the localization
length. Here we plot $\Delta_1$ as function of $1/\sqrt{\xi} \propto W^{1/(3-2K(U))}$. A linear
dependence on $1/\sqrt{\xi}$ with a slope depending on interaction $\Delta_0(U)$ is clear (lines).
As expected from interacting particles in a confining potential $\Delta_0(U) \propto U$
as can be seen in the inset}
\end{figure}

This behavior is seen in Fig. \ref{fig2} where $\Delta_1$ for different
values of disorder and interaction are indicated by the symbols.
Since (except for $W=0.3$) in all cases $\xi\le 200$) we have also averaged over
the different values of $L_A$ in the range $L/4<L_A<3L_A/4$, where the first
level spacing saturates. In the top figure, $\Delta_1$ is plotted as
function of $1/\sqrt{\xi_0} \propto W$. As indicated by the black line, for
the non-interacting case ($U=0$) as long as $\xi<L$ (i.e., $W > 0.4$) the numerical
data follows $1/\sqrt{\xi_0}$ perfectly. For the interacting cases, deviations are
clearly seen. That is not surprising since $\xi$ depends on $U$. Taking the dependence
of the localization length on the interactions into account by
$1/\sqrt{\xi} \propto W^{1/(3-2K(U))}$ a linear
relation of the form $\Delta_1=\Delta_0(U)/\sqrt{\xi}+{\rm Const}$
can be seen in the lower panel of Fig. \ref{fig2}. 
Moreover as can be seen in the inset $\Delta_0(U)$ depends linearly on $U$ as may be
expected from the level spacing of charged particles in a confined potential (Coulomb
blockade) \cite{averin86}. This further strengthens the picture of the entanglement spectrum
corresponding to a many-particle spectrum of an effective Hamiltonian with a 
confining potential near the boundary.

\begin{figure}
\includegraphics[width=8cm,height=!]{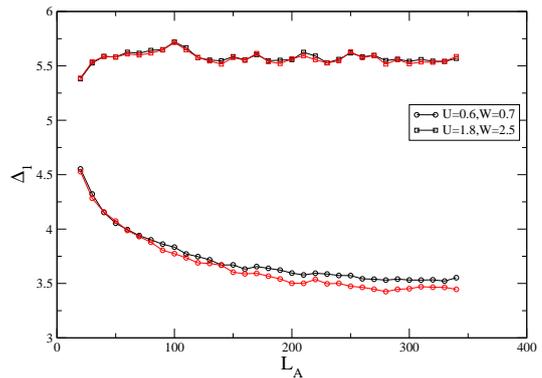}
\caption{\label{fig3}
Comparison of the averaged first entanglement level spacing between
the ground-state (black) and the first excited state (red). For the strongly
localized case $U=1.8,W=2.5$ there is no difference between the two. On the other
hand, for the weak disorder case, $U=0.6,W=0.7$, $\Delta_1$ is consistently 
lower in the saturated region for the excited state.
}
\end{figure}

Up to now we have considered the case for which the
pure state $|\Psi\rangle$ of the entire system
is the ground state. Of course, in principle one may calculate the entanglement
spectrum of the system for any pure state. Nevertheless, DMRG is suitable only for the
calculation of low-lying excitations, and therefore here we have calculated only
the entanglement spectrum for these states. In the strong disorder regime,
no physical difference is expected between the ground state and the low-lying 
excitations. Indeed, comparing $\Delta_1$ for the ground-state and the first excited
state for $U=1.8,W=2.5$ (Fig. \ref{fig3}), no significant difference
can be seen. On the other
hand, for weak disorder it has been show that even for low-lying excitations
there may be a significant increase in the localization length \cite{berkovits14}.
This can be seen for the weak disorder case of $U=0.6,W=0.7$, 
as a decrease in $\Delta_1$ for
the saturated area, as expected when $\xi$ increases.

\begin{figure}
\includegraphics[width=8cm,height=!]{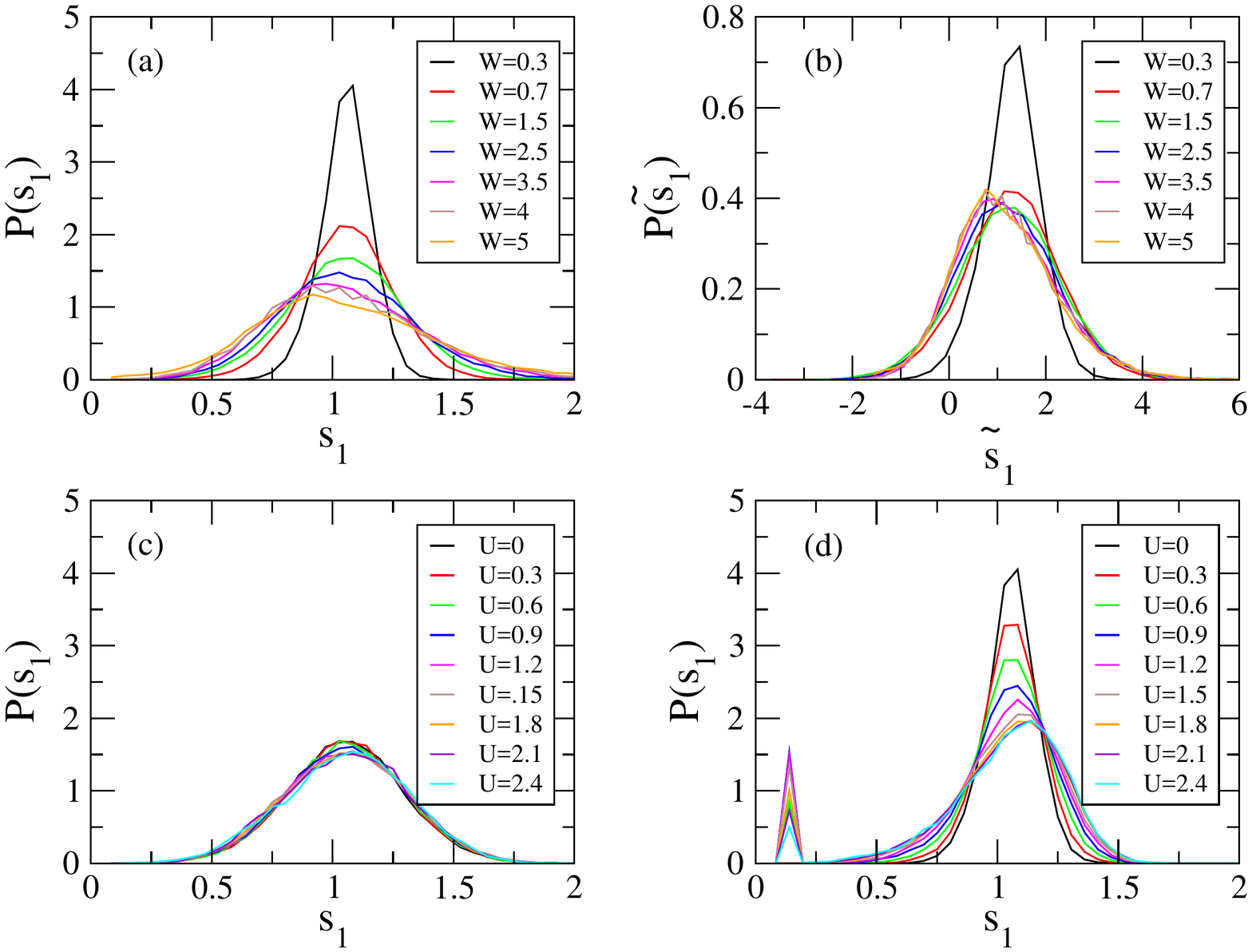}
\caption{\label{fig4}
  The numerical distribution of the normalized first entanglement level spacing $s_1$ for different
  strength of disorder $W$ and interaction strength $U$. (a) The distribution for different
  strength of disorder in the non-interacting ($U=0$) case. (b) As in (a) where
  $s_1$ is rescaled according to $\tilde s_1= (s_1-1)*(a+bW)+1$ and $P(\tilde s_1)= (a+bW)^{-1}$
  where $a$ and $b$ are constants. (c) The distribution for different
  strength of interaction strength at a given value of disorder $W=1.5$, the distribution is
  almost independent of $U$. (d) As in (c) for weaker disorder $W=0.7$. In this case as $U$ increases a second
peak in the distribution for small spacings appears.}
\end{figure}

\section{First entanglement level spacing distribution}
\label{s3}

One might naively expect that the distribution of the first entanglement level spacing
will be similar to the first excitation of a localized many-particle system, which
should follow the single-particle level spacing distribution, i.e., 
the Poisson distribution \cite{berkovits94}. The distribution of the
normalized first excitation $P(s_1)$ (where $s_1 = 
\varepsilon^{N_A}_2 - \varepsilon^{N_A}_1 )/ \Delta^{N_A}_1$)
is drawn from different realization of disorder, different cuts of $L_A$
in the range $L/4 <L_A < 3L/4$ and values of $N_A \sim L_A/2$ and presented
in Fig. \ref{fig4}(a) for the non-interacting ($U=0$) case. It is immediate
clear that this is not a Poisson distribution, but rather a
Gaussian-like broadening
of the spacing as function of $W$. This non-universal behavior of the 
distribution is due to the effective
confining potential of the entanglement Hamiltonian. Since only an
area of length $\xi$ is sampled by the entanglement spectrum the
system on this length scale is not localized. As is known for
disordered 1D systems
the distribution of single-particle level spacing crosses over very
rapidly from a universal Poisson distribution when
$\xi$ is smaller than the length to a non universal broadening as
$\xi$ becomes larger than the systems length, with no true
Wigner behavior in the  middle \cite{wobst02}.
Thus 
$P(s_1)$ shows the typical behavior of a short disordered 1D system.
The broadening of the distribution is proportional to $W$, and
the distribution might be rescaled by
$\tilde s_1= (s_1-1)*(a+bW)+1$ and $P(\tilde s_1)= (a+bW)^{-1}$
with the numerical factors $a=0.17$, $b=0.0375$. As can be seen in
\ref{fig4}(b)
after the rescalling the curves with stronger disorder
($W>0.7$) for which the localization length  is significantly
shorter than the system size, all curves fall on each other.

In the region where $\xi \ll L$ there is no dependence on $U$
as is demonstrated in \ref{fig4}(c) for the case of $W=1.5$.
Thus, in contrast with the average first level spacing which
depends on $\xi$ the distribution width depends only on the
on-site disorder $W$ and not on $U$ or $\xi$.
Nevertheless, for weak disorder ($W=0.7$, $\xi \geq L$) a peculiar
dependence on $U$ appears. As is seen in 
\ref{fig4}(d) a second peak in the distribution at low values
of $s_1$ appears. This peak has a non-monotonous
behavior as function of $U$. It is absent for $U=0$ increases
as $U$ increases up to $U=1.2$ and then decreases.
This feature is absent from stronger disordered samples (see, \ref{fig4}(c)).

\section{Intermediate regions in weakly disordered strongly interacting realizations}
\label{s4}

\begin{figure}
\includegraphics[width=8cm,height=!]{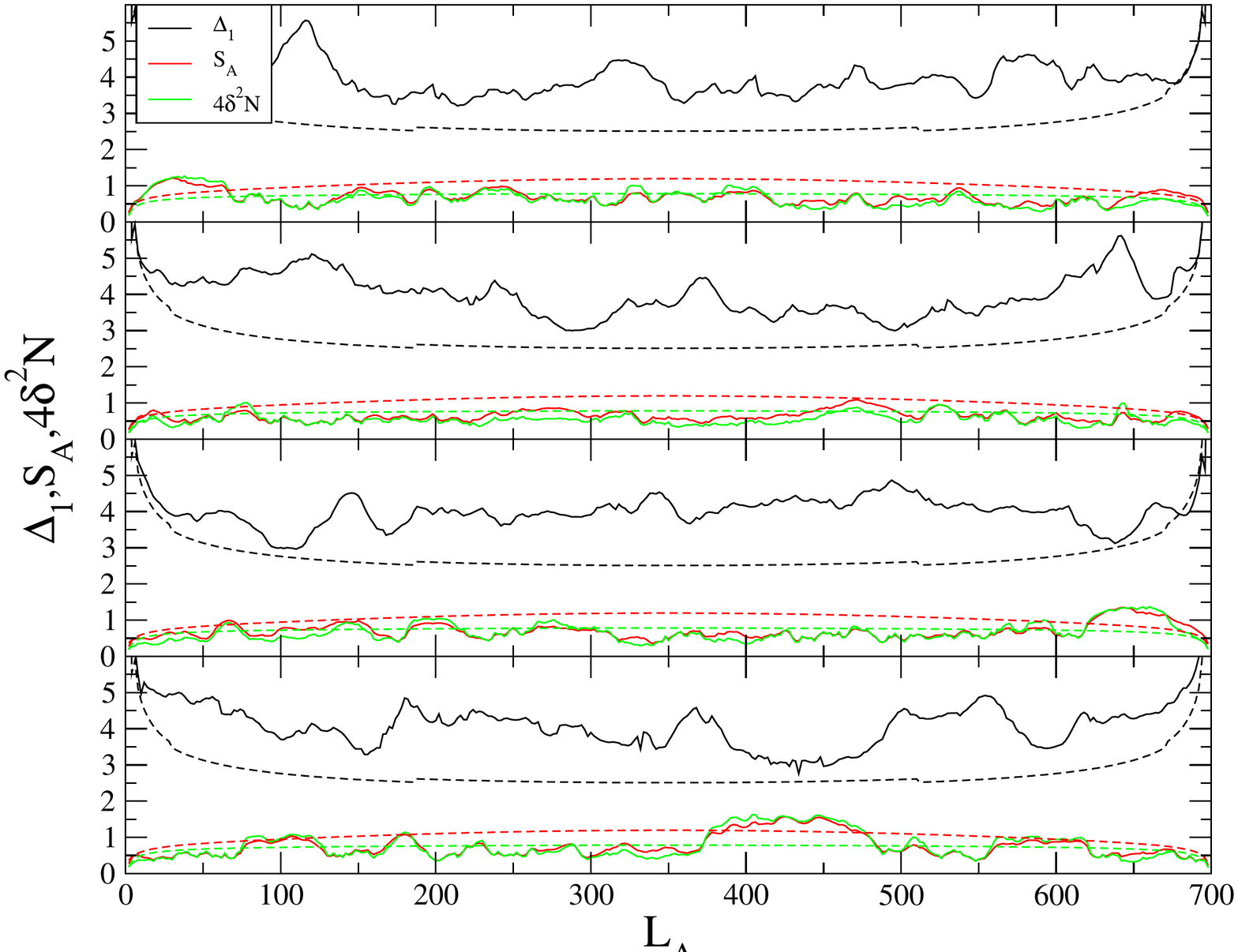}
\caption{\label{fig5}
  The first entanglement level spacing $s_1$ (black curve), the entanglement entropy $S_A$
  (red curve), and the
  particle number variance $\delta^2 N_A$ (green curve multiplied by 4 for clarity)
  for four different realizations of disorder,
  where $W=0.7$ and $U=2.4$ and $L_A=2,4,6,\ldots,L-2$.
  The results for a clean system ($W=0$, $U=2.4$) are presented
  by the corresponding dashed curves. 
}
\end{figure}

From where does this second peak for weakly disordered 
strongly interacting systems come from? 
Some insight may be gained from scrutinizing
specific realizations of disorder. Four representative realizations with
($W=0.7$, $U=2.4$) are shown in
Fig \ref{fig5}. For typical regions of each sample $s_1$ fluctuates around
the average and are significantly higher than for a clean case with the same interaction strength
($W=0$, $U=2.4$). Nevertheless,
there are rare regions (see, e.g., the lowest
panel of Fig \ref{fig5} in the region $370<L_A<470$)  where $s_1$ is
significantly lower then the average and much closer to the clean case value. 
Moreover, the entanglement entropy, $S_A$ is strongly enhanced in that region
even beyond the clean sample value. Likewise,
the particle number variance $\delta^2 N_A = \langle N_A^2 \rangle - \langle N_A\rangle^2 $
(easily calculated using DMRG),  is also enhanced in this region, much 
beyond to the values for a clean system ($W=0$) with the same interaction strength.
The correspondence between $S_A$ and $\delta^2 N_A$
seems to work well for these realizations although
strictly speaking, there is no theoretical proof for this relation in interacting
systems \cite{klich09,hsu09,song12a} . This will be further discussed elsewhere. Anyway, the behavior
of both $S_A$ and $\delta^2 N_A$ are in line with a ``metallic'' inclusion
within the localized sample. Thus, although as we have seen for $\Delta_1$,
on the average stronger interactions ($U$) corresponds to stronger localization
(smaller $\xi$),
there exist rare regions for  which the interplay between interaction and disorder
may lead to a more metallic-like behavior.
A similar behavior, where interactions lead to delocalization in rare realizations 
of 1D disordered systems, was
seen for the persistent current\cite{schmitteckert98}, and is also 
reminiscent of the intermediate “microemulsion” phases proposed for two-dimensional
systems \cite{spivak04}.

\begin{figure}
\includegraphics[width=8cm,height=!]{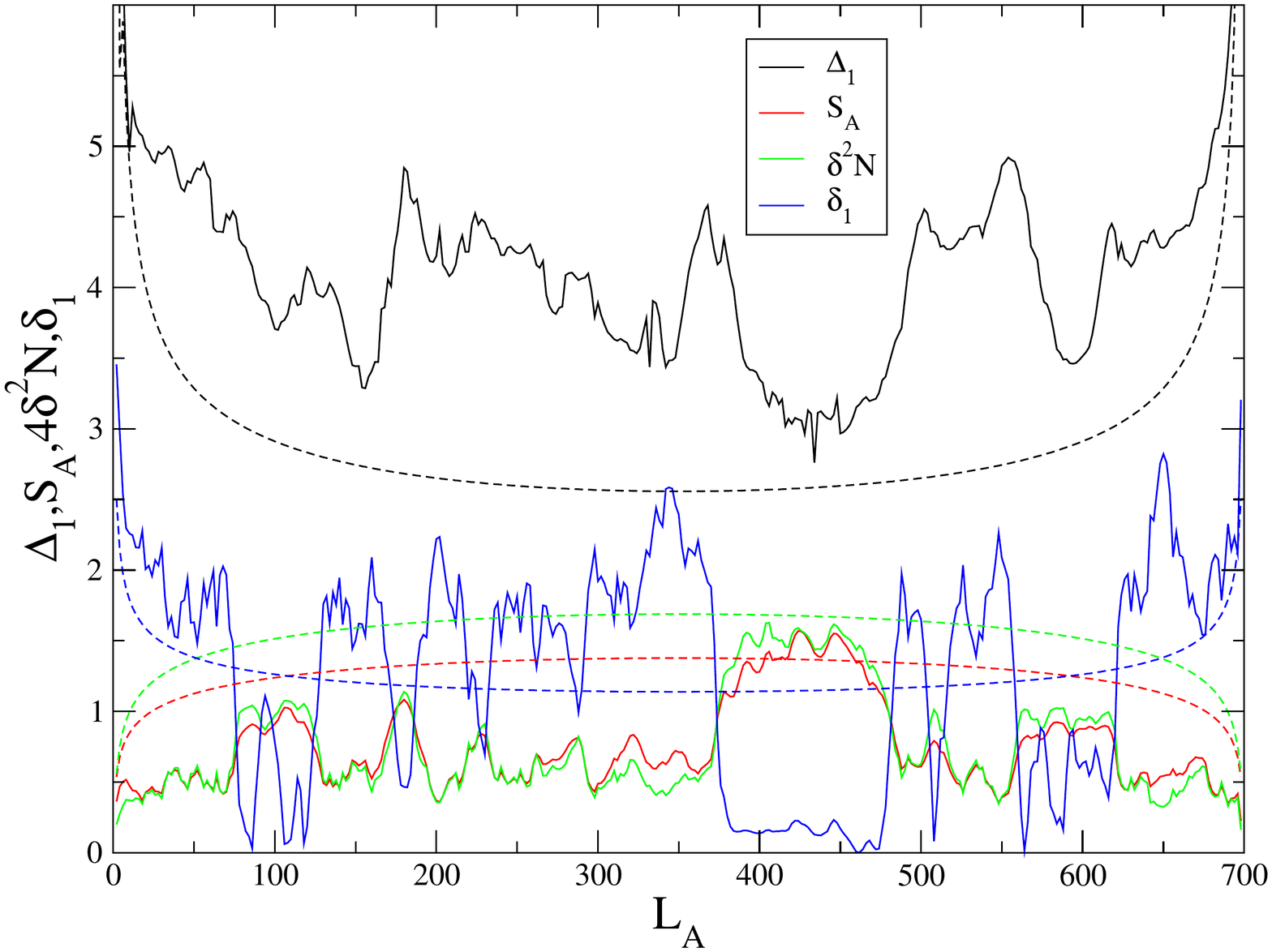}
\caption{\label{fig6}
  A closer look at the behavior of the last realization depicted in Fig. \ref{fig5}.
  The first entanglement level spacing $s_1$ (black curve), the entanglement entropy $S_A$
  (red curve), the
  particle number variance $\delta^2 N_A$ (green curve multiplied by 4 for clarity)
  and the spacing between two lowest entropy levels belonging to different particle number, $\delta_1$
  (blue curve). For comparison the same variables calculated for a clean non-interacting sample
  ($U=0$, $W=0$) are indicated by dashed curves.
}
\end{figure}

Some light on the nature of these rare regions can be shed by the low-lying  
entropy energies $\varepsilon^{N_A}_i$. Usually, for the half-filled case discussed here, and
a even partition ($L_A$ even), $\varepsilon^{N_A=L_A/2}_1$ is much lower than any other
energy, since a state with $N_A=L_A/2$ is the most probable. Thus, one expects 
$\delta_1=\min(\varepsilon^\{N_A=L_A/2 \pm 1_1\} - \varepsilon^{N_A=L_A/2}_1$ to be smaller
than $\Delta_1$, but not orders of magnitude lower. Indeed, comparing
$\delta_1$ to $\Delta_1$ for the last realization depicted in Fig. \ref{fig5}
(see Fig. \ref{fig6}) shows this behavior for most of the sample, except
for the region around $370<L_A<470$, and smaller regions around $L_A=100$ and
$L_A=600$ where $\delta_1$ reaches values close to zero. These are exactly the regions
where strongly enhanced values of $S_A$ and $\delta^2 N_A$ are seen, i.e.,
close to the values of a metallic sample ($U=0$, $W=0$).
For typical regions of the strongly interacting weakly disordered system
the ground state is a pinned charge density wave leading to small
variance in the number of particles and low entanglement entropy. 
In contrast, the rare regions are neither described by a charge density wave nor by a simple metallic behavior. This can
be clearly seen from the very different behavior of $\delta_1$ in these regions
compared to regular metals. As seen in Fig. \ref{fig6}, especially
for the region around $370<L_A<470$, $\delta_1$ is much lower for the disordered
rare region than for a clean system. This indicates that these rare regions are
governed by  different physics than the usual metallic 1D system.
The fact that $\delta_1 \sim 1$ indicates large and almost equal contribution
to the many-particle state from sectors with different number of particles in
region A, which is very different than for the clean metallic situation.

\section{Discussion}
\label{s5}

Thus, the average of the
first entanglement level spacing has been shown to have a clear relation to
the ground state
localization length and shows the expected behavior as function of the strength of the
on-site disorder and with the repulsive particle-particle interactions.
This stems from the fact that for a strongly localizaed system, the entanglement 
is confined to a distance of order of the localization
length from the boundary between the regions, and that has a clear imprint on
the low-lying eigenvalues of the RDM and the corresponding values of the entanglement
Hamiltonian. The distribution
of the first entanglement level spacing once the localization length is shorter
than the sample length is Gaussian-like and
quite universal. The distribution depends only on the strength
of disorder and not on the interactions. Such a behavior is actually expected
for the distribution of low-lying level spacings
in a disordered confining potential, as long as
the level spacing is larger than the influence of the disorder.  

On the other hand, for weakly disordered systems and strongly interacting systems,
the distribution shows
an interesting peak, signifying an increased probability for almost
degenerate level spacings. On closer examination of the behavior
for specific realization it becomes clear this feature is connected to
rare regions in the sample which exhibit metallic-like behavior. 
These rare regions in the ground state are composed of
more or less equal significant
contributions from two states with different number of particles. This
not only leaves a distinct signature in the entanglement spectrum, but also
leads to large variance in the number of particles in the region and high
entanglement of the order of the values seen for free fermions.
These intermediate “microemulsion” metallic phases embedded
in an insulating phase. Further study of their properties is needed as well
as their connection to the
phase separation in two-dimensional systems \cite{spivak04}, and to the
enhancement of the persistent current in rare disordered systems
\cite{schmitteckert98}.

\begin{acknowledgments}
Useful discussions with A. Turner and G. Murthy are gratefully acknowledged.
\end{acknowledgments}

\end{document}